\documentclass[aps,preprint,showpacs,floatfix]{revtex4-1}
\usepackage{bm}
\usepackage{amsmath}
\usepackage{epsfig}
\usepackage{rotating}
\usepackage{graphicx}

\def\pr{\prime}
\def\be{\begin{equation}}
\def\lan{\left\langle}
\def\ran{\right\rangle}
\def\ee{\end{equation}}
\def\barr{\begin{array}}
\def\earr{\end{array}}

\def\l{\left}
\def\r{\right}
\def\dis{\displaystyle}
\def\ed{\end{document}}
\def\f{\frac}

\def\bpi{{\mbox{\boldmath $\pi$}}}
\def\bnu{{\mbox{\boldmath $\nu$}}}

\def\cs{{\bf S}}

\def\ed{\end{document}}
\begin{document}

\title{Two species $k$-body embedded Gaussian unitary ensembles: $q$-normal form of the eigenvalue density}

\author{Manan Vyas}
\thanks{corresponding author, manan@icf.unam.mx}
\affiliation{Instituto de Ciencias F{\'i}sicas, Universidad Nacional Aut{\'o}noma de M\'{e}xico, 62210 Cuernavaca, M\'{e}xico}
\author{V. K. B. Kota}
\thanks{vkbkota@prl.res.in}
\affiliation{Physical Research  Laboratory, Ahmedabad 380 009, India}

\begin{abstract}

Eigenvalue density generated by embedded Gaussian unitary ensemble with $k$-body interactions for two species (say $\bpi$ and $\bnu$) fermion systems is investigated by deriving formulas for the lowest six moments. Assumed in constructing this ensemble, called EGUE($k:\bpi\bnu$), is that the $\bpi$ fermions ($m_1$ in number) occupy $N_1$ number of degenerate single particle (sp) states and similarly $\bnu$ fermions ($m_2$ in number) in $N_2$ number of degenerate sp states. The Hamiltonian is assumed to be $k$-body preserving $(m_1,m_2)$. Formulas with finite $(N_1,N_2)$ corrections and asymptotic limit formulas both show that the eigenvalue density takes $q$-normal form with the $q$ parameter defined by the fourth moment. The EGUE($k:\bpi\bnu$) formalism and results are extended to two species boson systems. Results in this work show that the $q$-normal form of the eigenvalue density established only recently for identical fermion and boson systems extends to two species fermion and boson systems.

\end{abstract}

%\pacs{}

\maketitle

\section{Introduction}

Investigating statistical properties of isolated finite many-particle quantum systems, such as atomic nuclei, mesoscopic systems (quantum dots, small metallic grains), ultra-cold atoms, quantum black holes with SYK model and so on, it is now well recognized that one needs random matrix models going beyond the classical Gaussian orthogonal (GOE) or unitary (GUE) or
symplectic (GSE) random matrix ensembles, by including fermion and boson nature of the particles, body-rank of the interactions and so on. One class of such random matrix ensembles that are studied in considerable detail in the last two decades are embedded ensembles (EE) in many fermion or boson spaces generated by two-body interactions (these ensembles are first introduced in \cite{FW,BF}). In the simplest version, the
$2$-particle Hamiltonian ($H$) of a spinless fermion (or boson) system is represented by GOE/GUE/GSE (all three classical ensembles combined are referred to as GE, Gaussian ensembles) and then the $m$ particle $H$ matrix is generated
using the $m$-particle Hilbert space geometry. As a GOE/GUE/GSE random matrix ensemble in $k$-particle spaces is embedded  in the $m$-particle $H$ matrix, these ensembles are generically called $2$-body embedded ensembles [EE($2$)]. For example, with GUE embedding we have Gaussian unitary ensemble of $2$-body interactions [EGUE(2)] and similarly EGOE(2). The EE($2$) ensembles are further extended to include the mean-field one-body part in the Hamiltonian giving EE($1+2$), various Lie algebraic symmetries such as $SU(2)$ generating spin, $SU(4)$ generating spin and isospin for atomic nuclei, parity etc. There is now large body of literature on various properties of these embedded ensembles with a variety of applications; see for example \cite{Br-81,Ko-01,Man-th,Ko-book,ILZ,KC-1,KC-2,IL-1,IL-2} and references therein. 
   
Going beyond EE with 2-body interactions, in the last 5 years in particular, the focus is shifted to EE($k$), i.e. EE generated by general $k$-body interactions. For very early studies of EE($k$) see \cite{MF,Br-81,FW,BRW1,BRW2,ABRW,Ko-05,Volya,Small1,Small2} and similarly for more recent studies see \cite{MB,Bent1,Bent2,Bent3,MK-1,MK-2,MK-3,MK-4,CH1,CH2,CH3,Ko-arx}. One significant property of 
EE($k$) is that the eigenvalue density of these ensembles with $m$ identical spinless fermions or bosons occupying say $N$ single particle (sp) states, changes from Gaussian form (valid for $k << m)$ to semi-circle form (valid for $k=m$). This feature was established very early by Mon and French \cite{MF}  by deriving asymptotic limit formulas (say $N \rightarrow \infty$, $m \rightarrow \infty$, $m/N \rightarrow 0$ and $k$ finite) for lower order moments of the eigenvalue density. However, only recently it is recognized that the eigenvalue density in fact takes $q$-normal form for a general $k$ and the $q$-normal goes to Gaussian for $k/m \rightarrow 0$ (or $q \rightarrow 1$) and semi-circle for $k=m$ ($q=0$). The $q$-normal is formally established, via lower order moments, for EGUE($k$) for spinless fermion systems with identical fermions \cite{MK-1} and seen to extend to spinless boson systems with identical bosons \cite{MK-1,CH1}. Also, formulas for the $q$ parameter are derived in \cite{MK-1} for EGOE($k$) and EGUE($k$)
for these systems. The purpose of the present paper is to go beyond identical spinless fermion/boson systems and consider EE(k) for two species fermion and boson systems. In this paper, we will restrict all the discussion to EGUE($k$) for two species (spinless) fermion and boson systems [called EGUE($k:\bpi\bnu$) and BEGUE($k:\bpi\bnu$) respectively] and derive formulas for the lower order moments of the eigenvalue density generated by these ensembles to establish that for these extended systems also, the 
$q$-normal form applies.  Studying moments of eigenvalue density is needed for applying statistical nuclear spectroscopy based on $q$-normal forms as developed in \cite{MK-4, KM-15}.   As we are addressing only the form of the smoothed eigenvalue density (ignoring fluctuations), it is well known that the smoothed form needs often only the first four moments; eigenfunction structure or equivalently transition strengths distributions will be considered elsewhere.  

We consider a system with two types of spinless fermions (or bosons) $\bpi$ and $\bnu$.  Say this system consists of $m_1$ number of $\bpi$ fermions (or bosons) in $N_1$ number of single particle (sp) states and $m_2$ number of $\bnu$ fermions (or bosons) in $N_2$ number of sp states. This is similar to protons ($\bpi$) and neutrons ($\bnu$) in an atomic nucleus. (or a two species bosonic system).  Further, the system Hamiltonian ($H$) operator is assumed to be $k$-body preserving $(m_1,m_2)$.  Thus, we have two spaces with the first one being (\#1) or $\bpi$  and the second one being (\#2) or $\bnu$ space.  If we consider strong mixing between the two spaces, with $m = m_1 + m_2$ fixed,  then we are back to the simple EGUE($k$) considered in \cite{MK-1} giving again $q$-normal form for eigenvalue density. However,  if the mixing is weak,  then we have a new ensemble and this ensemble may be studied in future. Now, we will give a preview.

In Section \ref{sec1},  for completeness, definition of the EGUE($k:\bpi\bnu$) ensemble is presented and also given is the $q$-normal form. In addition, for use in Sections \ref{sec3}-\ref{sec5}, some important results from $U(N)$ algebra for EGUE($k$) are presented. In Section \ref{sec3}, formulas are derived for the second, fourth and sixth moment of the eigenvalue density generated by  EGUE($k:\bpi\bnu$) ensemble using the $U(N)$ Wigner-Racah algebra. In Section \ref{sec4} asymptotic (or dilute limit) formulas for the second, fourth and sixth moment are presented. In addition, numerical results using several examples are presented for the $q$ parameter (defined by the fourth moment) and for the sixth moment. Results for the sixth moment establish the $q$-normal form for the eigenvalue density generated by  EGUE($k:\bpi\bnu$) ensemble. This result is assumed to be true in Ref.  \cite{MK-4} where Statistical Nuclear Spectroscopy with $q$-normal form is developed. Section \ref{sec5} presents results for the bosonic BEGUE($k:\bpi\bnu$) ensemble.  Finally, Section \ref{sec6} gives conclusions.

\section{Preliminaries}
\label{sec1}

\subsection{Definition of EGUE($k:\bpi\bnu$) for two species fermion systems}

Let us consider a system with two types of spinless fermions $\bpi$ and $\bnu$. Say this system consists of $m_1$ number of $\bpi$ fermions in $N_1$ number of single particle (sp) states and $m_2$ number of $\bnu$ fermions in $N_2$ number of sp states. This is
similar to protons ($\bpi$) and neutrons ($\bnu$) in an atomic nucleus. Further, the system Hamiltonian ($H$) operator is assumed to be $k$-body preserving $(m_1,m_2)$. Thus, we have two spaces with the first one (\#1) or $\bpi$ space having $N_1$ sp states and the second (\#2) or $\bnu$ space $N_2$ sp states. States of a $(m_1,m_2)$ system  can be represented by$\l.\l|m_1,\gamma : m_2, x\r.\ran$ where the Greek label $\gamma$ gives the additional label needed for specifying completely the many particle states generated by $\bpi$ fermions in the first space and similarly, the Roman label $x$ gives the additional label needed for specifying completely the many particle states generated by $\bnu$ fermions in the second space. The $H$ matrix dimension in $(m_1,m_2)$ space is $d(m_1,m_2) =\binom{N_1}{m_1}\,\binom{N_2}{m_2}$. Proceeding further, the $k$-body $H$ operator preserving $(m_1,m_2)$ is given by,
\be
\barr{l}
H(k) = \dis\sum_{i+j=k} \dis\sum_{\alpha , \beta \in i} \dis\sum_{a,b \in j}\;
V_{\alpha a:\beta b}(i,j)\;
A^\dagger_{\alpha}(i)\, A_{\beta}(i)\, A^\dagger_a(j)\,A_b(j)\;; \\ \\
V_{\alpha a:\beta b}(i,j) = \lan i, \alpha : j, a \mid H \mid i, \beta : 
j,b\ran\;.
\earr \label{eq.pn4}
\ee
Note that $A^\dagger_{\alpha}(i)$  creates the normalized state
$\l.\l|i,\alpha\r.\ran$ with $i$ number of $\bpi$ fermions and
$A^\dagger_a(j)$ creates the normalized state $\l.\l|j,a\r.\ran$ with $j$ number of $\bnu$
fermions. Similar is the action of the annihilation
operators $A_{\beta}(i)$ and $A_b(j)$. For example for a two-body Hamiltonian,
$(i,j)=(2,0)$, $(1,1)$ and $(0,2)$ and similarly, for a 3-body Hamiltonian  $(i,j)=(3,0)$, $(2,1)$, $(1,2)$ and $(0,3)$ an so on. For each $(i,j)$ pair with $i+j=k$, we have a matrix $V(i,j)$ in the
$k$-particle space and the $H$ matrix in $k$ particle space is a direct sum of these  matrices. Their dimensions being $\binom{N_1}{i} \binom{N_2}{j}$. Action of the $H$ operator on the states of a $(m_1,m_2)$ system  generates the $H$ matrix in the $(m_1,m_2)$ space.
Given all these, we will now introduce the EGUE($k$) ensemble for the $\bpi-\bnu$ system, i.e. EGUE($k:\bpi\bnu$).
Firstly, the $V(i,j)$ matrices in Eq. (\ref{eq.pn1}) are represented by independent GUEs with matrix elements being zero centered Gaussian variables and variances satisfying,
\be
\overline{V_{\alpha a : \beta b}(i,j)\;V_{\alpha^\pr a^\pr : \beta^\pr b^\pr}
(i^\pr ,j^\pr)} 
=v^2(i,j)\;\delta_{i i^\pr} \; \delta_{j j^\pr} \; \delta_{\alpha \beta^\pr} \;
\delta_{a b^\pr} \; \delta_{\beta \alpha^\pr} \; \delta_{b a^\pr} \;.
\label{eq.pn5}
\ee
The $v^2(i,j)$ here are free parameters that depend on $(i,j)$. 
With this independent GUE representation of the $V$ matrices, the action of the $H$ operator in Eq. (\ref{eq.pn4}), for a given member of the $V$ ensemble, on the states in $(m_1,m_2)$ space will generate a $H$ matrix of dimension $\binom{N_1}{m_1}\,\binom{N_2}{m_2}$. Repeating this for all the $V$ matrix members, we will have an ensemble of random matrices in $(m_1,m_2)$ space and this is the EGUE($k:\bpi\bnu$) ensemble. It is important to note that the embedding algebra for the EGUE($k:\bpi\bnu$)  generated by the action of the $V$ ensemble on $\l.\l|m_1, \alpha : m_2, a\r.\ran$ states is the direct sum algebra  $U_{\bpi}(N_1) \oplus U_{\bnu}(N_2)$. Thus, in group theory language, the EGUE($k:\bpi\bnu$) ensemble is in fact EGUE($k$)-$[U_{\bpi}(N_1) \oplus U_{\bnu}(N_2)]$ ensemble. Our interest in the present paper is to establish that the form of the eigenvalue density for EGUE($k:\bpi\bnu$) is $q$-normal and to this end we will derive formulas for the lowest six moments of the eigenvalue density in Section 3 using $U(N)$ algebra as given in \cite{Ko-05} (see also \cite{BRW1}). Before going further, let us mention that EGUE($k:\bpi\bnu$) was discussed
earlier in \cite{KM-15} and there it is assumed that the eigenvalue density generated by this ensemble is a Gaussian. 

\subsection{$q$-normal and $q$-Hermite polynomials}

In order to introduce the $q$-normal form, firstly one needs the
definition of $q$ numbers $[n]_q$ and they are(with $[0]_q=0$), 
\be
\l[n\r]_q = \dis\frac{1-q^n}{1-q} = 1+q + q^2 + \ldots+q^{n-1}\;.
\label{eq.q1}
\ee
Note that $[n]_{q \rightarrow1}=n$. Similarly, $q$-factorial $[n]_q! = \dis\Pi^{n}_{j=1} \,[j]_q$ with $[0]_q!=1$. Given these, the $q$-normal distribution $f_{qN}(x|q)$ with $x$ being a standardized variable (then $x$ is zero centered with variance unity), is given by \cite{Ismail,Sza-1} 
\be
f_{qN}(x|q) = \dis\frac{\dis\sqrt{1-q} \dis\prod_{k^\pr=0}^{\infty} \l(1-
q^{k^\pr +1}\r)}{2\pi\,\dis\sqrt{4-(1-q)x^2}}\; \dis\prod_{k^\pr=0}^{\infty}
\l[(1+q^{k^\pr})^2 - (1-q) q^{k^\pr} x^2\r]\;.
\label{eq.q2}
\ee
The $f_{qN}(x|q)$ is non-zero for $x$ in the domain defined by $\cs(q)$ where
\be
\cs(q) = \l(-\dis\frac{2}{\dis\sqrt{1-q}}\;,\;+\dis\frac{2}{\dis\sqrt{1-q}}\r)\;.
\label{eq.q3}
\ee
Note that $\int_{\cs(q)} f_{qN}(x|q)\,dx =1$. Most important property of $q$-normal is that $f_{qN}(x|1)$ is Gaussian with $\cs(q=1)=(-\infty , \infty)$ and similarly, $f_{qN}(x|0)=(1/2\pi) \sqrt{4-x^2}$, the semi-circle with $\cs(q=0)=(-2,2)$. Also, the the reduced fourth moment $\mu_4$ and the reduced sixth moment $\mu_6$ of $f$ are, 
\be
\mu_4=2+q\;,\;\;\;\;\mu_6=5+6q+3q^2+q^3\;.
\label{eq.q3a}
\ee
Thus, it is possible to use $\mu_4$ to determine the $q$ parameter and test the goodness of $q$-normal form by verifying $\mu_6$ if this follows the formula in Eq. (\ref{eq.q3a}); see Sections 3-6 ahead. Going further, The $q$-Hermite polynomials $He_n(x|q)$, that are orthogonal with $f_{qN}$ as the weight function, are defined by the recursion relation
\be
x\,He_n(x|q) = He_{n+1}(x|q) + \l[n\r]_q\,He_{n-1}(x|q)
\label{eq.q4}
\ee
with $He_0(x|q)=1$ and $He_{-1}(x|q)=0$. Note that for $q=1$, the $q$-Hermite polynomials reduce to normal Hermite polynomials (related to Gaussian) and for $q=0$ they will reduce to Chebyshev polynomials (related to semi-circle). For example, $He_0(x|q) = 1$,\;, $He_1(x|q) = x$, $He_2(x|q) = x^2-1$, $He_3(x|q) =  x^3-(2+q)x$ and $He_4(x|q) = x^4-(3+2q+q^2)x^2+(1+q+q^2)$.

\subsection{Basic results from $U(N)$ algebra for EGUE($k$)}

In this subsection we restrict our discussion to EGUE($k$) for a system of $m$ identical spinless fermions in $N$ single particle states. Then, all the $m$-fermion states belong to the totally antisymmetric irreducible representation (irrep) $f_m=\{1^m\}$ of $U(N)$ (note that we are using Young tableaux notation for irreps). It is useful to note that the irrep conjugate to the $\{1^m\}$ irrep is $\overline{f_m}=\{1^{N-m}\}$(the 'bar' used here for denoting conjugate irrep should not be confused with the 'bar' that will be used later for ensemble averages) . The EGUE($k$) is generated by $k$-body Hamiltonians $H(k)$ that form GUE in $k$-particle space (with matrix elements variance unity)and the $H(k)$ operator can be decomposed into $U(N)$ tensors $B^\nu(k)$ with the irreps $\nu = \{2^\nu 1^{N-2\nu}\}$ and $\nu=0,1,\ldots,k$; note that $\nu = \overline{\nu}$ . Also, $\nu =0$ corresponds to $\{1^N\} = \{0\}$ irrep; see \cite{Ko-05}. Given $m$ fermion states $\l.\l|f_m,\alpha\r.\ran$ where $\alpha$ are additional labels needed for complete specification of the $m$ fermion state, we denote the $H$ matrix elements in $m$ fermion space by 
$$
H_{\alpha_1 \alpha_2}=\lan f_m \alpha_1 \mid H(k) \mid f_m 
\alpha_2\ran\;.
$$
Now, two important results, for the ensemble average of the product any two $m$-particle matrix elements of $H$, that follow from the $U(N)$ Wigner-Racah algebra are \cite{Ko-05,BRW1},
\be
\barr{rcl}
\overline{H_{\alpha_1 \alpha_2}H_{\alpha_3 \alpha_4}} & = &
\dis\sum_{\nu = 0,1,\ldots,k; \omega_\nu} \Lambda^\nu(N,m,m-k)\;C^{\nu , \omega_{\nu}}_{\alpha_1\;\overline{\alpha_2}}\;C^{\nu , \omega_{\nu}}_{\alpha_3\;\overline{\alpha_4}}\;,
\earr \label{eq.pn1}
\ee
and
\be
\overline{H_{\alpha_1 \alpha_2}H_{\alpha_3 \alpha_4}} = \dis\sum_{\mu=0,1,\ldots,m-k; \omega_\mu} \Lambda^\mu(N,m,k) \;C^{\mu , \omega_{\mu}}_{\alpha_1\;\overline{\alpha_4}}\;C^{\mu , \omega_{\mu}}_{\alpha_3\;\overline{\alpha_2}}
\label{eq.pn2}
\ee
with
\be
\Lambda^{\nu}(N,m,r) = \binom{m-\nu}{r}\,\binom{N-m+r-\nu}{r}\;.
\label{eq.pn3}
\ee
Here, $C^{\nu , \omega_\nu}_{\alpha_1\;\overline{\alpha_2}}$ are
$SU(N)$ Clebsh-Gordan (CG) coefficients $\lan f_m \alpha_1 \;\;\overline{f_m}\,\overline{\alpha_2} \mid \nu\; \omega_\nu\ran$. Just as the $\alpha's$, the $\omega_\nu$ are additional labels that correspond to a given $\nu$.

\section{Formulas for second, fourth and sixth moment with finite $(N_1,N_2)$ for EGUE($k:\bpi\bnu$)}
\label{sec3}

Moments of the ensemble averaged eigenvalue density generated by
EGUE($k:\bpi\bnu$) are given by $\overline{\lan H^p\ran^{m_1,m_2}}$ with $p=1,2,\ldots$; note that the over-line denotes ensemble average and the notation ${\lan H^p\ran^{m_1,m_2}}$ means the trace of $H^p$ over the space defined by $(m_1,m_2)$ divided by the dimension of the $(m_1,m_2)$ space.  By definition of EGUE($k:\bpi\bnu$), it is easy to see that all the odd moments ($p$ odd) are zero. Lowest even moment with $p=2$ defines the scale and the forth moment $p=4$ defines the $q$ parameter of the $q$-normal (see Section 2.2). Thus, for minimal verification of the $q$-normal form, we need the sixth ($p=6$) moment. Formulas for these are derived in this Section. As formulas for $p=2$ and $4$ are already given in \cite{KM-15}, their derivation is presented here for completeness and also for introducing some notations.

Firstly, from Eqs. (\ref{eq.pn4}) and (\ref{eq.pn5}) it is easy to see that the ensemble averaged moments $\overline{\lan H^p\ran^{m_1,m_2}}$ involve products of $v^2(i,j)$.  As mentioned before,  $v^2(i,j)$'s depend not only on $i$ and $j$ but also on parameter $k$. Therefore, we can derive formulas for the moments by assuming $v^2(i,j)=1$ and later put back the $v^2(i,j)$'s appropriately.   With this we can write $H$ as,
\be
H(k) = \dis\sum_{i+j=k} H(i,j) = \dis\sum_{i+j=k} H_1(i) H_2(j)\;.
\label{eq.pn6}
\ee
Here, $H_1(i)$ is $i$-body operator in $\bpi$ space and similarly $H_2(j)$ is $j$-body in $\bnu$ space. In addition we use the so-called binary correlation approximation described and used in several papers in the past; see for example \cite{MF,Br-81,Ko-book,BRW1,Ko-05,Small1,KM-15}. With these, we have easily for the second moment,
\be
\barr{rcl}
\overline{\lan H^2\ran^{m_1,m_2}} & = & \dis\sum_{i+j=k} \overline{\lan H(i,j) H(i,j)\ran^{m_1,m_2}} \\ 
& = & \dis\sum_{i+j=k} v^2(i,j)\; \overline{\lan H^2_1(i)\ran^{m_1}}\;\;\overline{\lan H^2_2(j)\ran^{m_2}}\;.
\earr \label{eq.pn7}
\ee
Here, in the second step we have used Eq. (\ref{eq.pn5}). Then, Eq. (\ref{eq.pn2}) gives easily the formula \cite{Ko-05,BRW1},
\be
\overline{\lan H^2\ran^{m_1,m_2}} = \dis\sum_{i+j=k} v^2(i,j)\; \Lambda^0(N_1,m_1,i) \,\Lambda^0(N_2,m_2,j)\;.
\label{eq.pn7a}
\ee
Going further and using binary correlation approximation we have for the fourth moment three terms (two irreducible terms),
\be
\barr{rcl}
\overline{\lan H^4\ran^{m_1,m_2}} & = & 
\overline{\lan AABB\ran^{m_1m_2}} + \overline{\lan ABBA\ran^{m_1m_2}} + \overline{\lan ABAB\ran^{m_1m_2}} \\
& = &
2\l[\overline{\lan H^2\ran^{m_1,m_2}}\r]^2 + \overline{\lan ABAB\ran^{m_1m_2}}\;.
\earr \label{eq.pn8}
\ee
Note that the operator $A$, $B$ etc.  are $H$ operators and the $H$'s with the same symbol ($A$, $B$ etc.) are correlated $H$'s. With this the first two terms in the second step above simplifies to the first term in the last step. The last term is non-trivial and this is simplified as follows,
\be
\barr{l} 
\overline{\lan ABAB\ran^{m_1m_2}} =
\dis\sum_{i+j=k,p+q=k} \overline{\lan A(i,j)B(p,q)A(i,j)B(p,q)\ran^{m_1,m_2}} \\
= \dis\sum_{i+j=k,p+q=k} v^2(i,j)v^2(p,q)\;\overline{\lan A_1(i) B_1(p) A_1(i)B_1(p)\ran^{m_1}}\;\;\;\overline{\lan A_2(j) B_2(q) A_2(j)B_2(q)\ran^{m_2}}
\earr \label{eq.pn9}
\ee 
Here again, $A_1(i)$'s are correlated and similarly $B_1(p)$'s, $A_2(j)$'s and $B_2(q)$'s. Then, for example with $A_{1;\alpha_1 \alpha_2}=\lan m_1 \alpha_1 \mid H_1 \mid m_1 \alpha_2\ran$, we have by inserting intermediate states between the $H$'s in $\lan H^4\ran$,  
\be  
\barr{l}
\overline{\lan A_1(i) B_1(p) A_1(i) B_1(p)\ran^{m_1}} = \dis\f{1}{\binom{N_1}{m_1}} \dis\sum_{\alpha_1,\alpha_2,\alpha_3,\alpha_4}
\overline{H_{1;\alpha_1 \alpha_2}(i)\,H_{1;\alpha_3 \alpha_4}(i)}\;\;\overline{H_{1;\alpha_2 \alpha_3}(p)\,H_{1;\alpha_4 \alpha_1}(p)} \\
= \dis\f{1}{\binom{N_1}{m_1}}\;\dis\sum_{\nu=0}^{min(i,m_1-p)}
\Lambda^\nu(N_1,m_1,m_1-i)\,\Lambda^\nu(N_1,m_1,p)\, d(N_1:\nu)\;.
\earr \label{eq.pn10}
\ee
Here we used Eqs. (\ref{eq.pn1}) and (\ref{eq.pn2}) and the sum rules for the CG coefficients; see \cite{Ko-05} for details.  The $\Lambda$'s in Eq. (\ref{eq.pn10}) are defined by Eq. (\ref{eq.pn3}) and $d(N:\nu) = \binom{N}{\nu}^2 - \binom{N}{\nu-1}^2$. Combining Eqs. (\ref{eq.pn8}), (\ref{eq.pn9}) and (\ref{eq.pn10}) finally give for the fourth moment the formula,
\be
\barr{rcl}
\overline{\lan H^4\ran^{m_1,m_2}} & = &  
2\l[\overline{\lan H^2\ran^{m_1,m_2}}\r]^2 + \overline{\lan ABAB\ran^{m_1m_2}}\;;\\
\overline{\lan ABAB\ran^{m_1m_2}}& = & \dis\sum_{i+j=k,p+q=k}
v^2(i,j)\,v^2(p,q)\;Z(N_1,m_1,i,p)\, Z(N_2,m_2,j,q)
\earr \label{eq.pn11}
\ee
where
\be
Z(N,m,k_1,k_2) = \dis\f{1}{\binom{N}{m}}\;\dis\sum_{\nu=0}^{min(k_1,m-k_2)} \Lambda^\nu(N,m,m-k_1)\,\Lambda^\nu(N,m,k_2) \,d(N:\nu)\;.
\label{eq.pn11a}
\ee
Now we will derive the formulas for the sixth moment.

Using binary correlation approximation, we have for the sixth moment fifteen terms and they will simplify to four irreducible terms giving,
\be
\barr{l}
\overline{\lan H^6\ran^{m_1,m_2}} =
5\overline{\lan AABBCC\ran^{m_1m_2}} + 6\overline{\lan AABCBC\ran^{m_1m_2}} + 3\overline{\lan ABACBC\ran^{m_1m_2}} +
\overline{\lan ABCABC\ran^{m_1m_2}} \\
= 5 \l[\overline{\lan H^2\ran^{m_1,m_2}}\r]^3 + 6 \l[\overline{\lan H^2\ran^{m_1,m_2}}\r]\; \overline{\lan BCBC\ran^{m_1m_2}} + 3\overline{\lan ABACBC\ran^{m_1m_2}} +
\overline{\lan ABCABC\ran^{m_1m_2}} \;.
\earr \label{eq.pn12}
\ee 
Simplification of the last two terms here are as follows. Firstly,
$\overline{\lan ABACBC\ran^{m_1m_2}}$ is,
\be
\barr{l}
\overline{\lan ABACBC\ran^{m_1m_2}} = 
\dis\sum_{i+j=k,p+q=k,r+s=k} \overline{\lan A(i,j)B(p,q)A(i,j)C(r,s)B(p,q)C(r,s)\ran^{m_1,m_2}} \\
= \dis\sum_{i+j=k,p+q=k,r+s=k} v^2(i,j)\,v^2(p,q)\,v^2(r,s)\;\overline{\lan A_1(i) B_1(p) A_1(i) C_1(r) B_1(p) C_1(r)\ran^{m_1}} \\
\times \; \overline{\lan A_2(j) B_2(q) A_2(j) C_2(r) B_2(q) C_2(r)\ran^{m_2}}\;.
\earr \label{eq.pn13}
\ee 
Now, for example
\be  
\barr{l}
\overline{\lan A_1(i) B_1(p) A_1(i) C_1(r) B_1(p) C_1(r)\ran^{m_1}} = \\
\dis\f{1}{\binom{N_1}{m_1}} \dis\sum_{\alpha_1,\alpha_2,\alpha_3,\alpha_4,\alpha_5,\alpha_6}
\overline{H_{1;\alpha_1 \alpha_2}(i)\,H_{1;\alpha_3 \alpha_4}(i)}\;\;\overline{H_{1;\alpha_2 \alpha_3}(p)\,H_{1;\alpha_5 \alpha_6}(p)}\;\; \overline{H_{1;\alpha_4 \alpha_5}(r)\,H_{1;\alpha_6 \alpha_1}(r)}\;\; \\
= \dis\f{1}{\binom{N_1}{m_1}}\;\dis\sum_{\nu=0}^{min(p,m_1-i,m_1-r)}
\Lambda^\nu(N_1,m_1,i)\,\Lambda^\nu(N_1,m_1,m_1-p)\, \Lambda^\nu(N_1,m_1,r)\, d(N_1:\nu)\;.
\earr \label{eq.pn14}
\ee
Here, to obtain the last equality, again we have used Eqs. (\ref{eq.pn1}) and (\ref{eq.pn2}) and applied the sum rules for the CG coefficients. Combining Eqs. (\ref{eq.pn13}) and(\ref{eq.pn14}) will give the formula for $\overline{\lan ABACBC\ran^{m_1,m_2}}$. Finally, we need the formula for
$\overline{\lan ABCABC\ran^{m_1m_2}}$ and in the first step we have,
\be
\barr{l}
\overline{\lan ABCABC\ran^{m_1m_2}} = 
\dis\sum_{i+j=k,p+q=k,r+s=k} \overline{\lan A(i,j)B(p,q)C(r,s)A(i,j)B(p,q)C(r,s)\ran^{m_1,m_2}} \\
= \dis\sum_{i+j=k,p+q=k,r+s=k} v^2(i,j)\,v^2(p,q)\,v^2(r,s)\;\overline{\lan A_1(i) B_1(p) C_1(r) A_1(i) B_1(p) C_1(r)\ran^{m_1}} \\
\times \;\overline{\lan A_2(j) B_2(q) C_2(r) A_2(j) B_2(q) C_2(r)\ran^{m_2}}\;.
\earr \label{eq.pn15}
\ee 
Proceeding further, we can write $\lan ----- \ran^{m_1}$ as,
\be
\barr{l}  
\overline{\lan A_1(i) B_1(p) C_1(r) A_1(i) B_1(p) C_1(r)\ran^{m_1}} = \dis\f{1}{\binom{N_1}{m_1}} \dis\sum_{\alpha_1,\alpha_2,\alpha_3,\alpha_4,\alpha_5,\alpha_6} \\
\overline{H_{1;\alpha_1 \alpha_2}(i)\,H_{1;\alpha_4 \alpha_5}(i)}\;\;\overline{H_{1;\alpha_2 \alpha_3}(p)\,H_{1;\alpha_5 \alpha_6}(p)}\;\; \overline{H_{1;\alpha_3 \alpha_4}(r)\,H_{1;\alpha_6 \alpha_1}(r)}\;.
\earr \label{eq.pn16}
\ee
For further simplification we need to use not only Eq. (\ref{eq.pn1}) and (\ref{eq.pn2}) but also Eqs. (32)-(36) of \cite{KM-15} (see also \cite{Ko-arx}). With these, one can see that simplification of Eq. (\ref{eq.pn16}) needs a $SU(N)$ Racah coefficient. Without going into details, then we have the formula
\be
\barr{l}
\overline{\lan A_1(i) B_1(p) C_1(r) A_1(i) B_1(p) C_1(r)\ran^{m_1}} = \dis\f{1}{\binom{N_1}{m_1}^2} \dis\sum_{\nu_1=0}^{i} \dis\sum_{\nu2=0}^{p} \dis\sum_{\nu_3=0}^{m_1-r}  d(N_1:\nu_1) \,d(N_1:\nu_2) \\
\times \; \Lambda^{\nu_1}(N_1,m_1,m_1-i)\,\Lambda^{\nu_2}(N_1,m_1,m_1-p)\, \Lambda^{\nu_3}(N_1,m_1,r)\,\l|U(f_m \nu_1 f_m \nu_2\,;\, f_m \nu_3)\r|^2\;.
\earr \label{eq.pn17}
\ee
Here, $U(f_m \nu_1 f_m \nu_2\,;\, f_m \nu_3)$ is a $SU(N)$ Racah coefficient; for the definition and properties of $SU(N)$ Racah coefficients see for example \cite{Butler1,Butler2}. Combining Eqs. (\ref{eq.pn12})-(\ref{eq.pn17}) give the final formula for
the sixth moment, 
\be
\barr{l}
\overline{\lan H^6\ran^{m_1,m_2}} = 5\, \l[\overline{\lan H^2\ran^{m_1,m_2}}\r]^3 \\
+ 6\, \l[\overline{\lan H^2\ran^{m_1,m_2}}\r]\; \dis\sum_{i+j=k,p+q=k} v^2(i,j)\,v^2(p,q)\;Z(N_1,m_1,i,p)\, 
Z(N_2,m_2,j,q) \\
+ 3\,\dis\sum_{i+j=k,p+q=k,r+s=k} v^2(i,j)\,v^2(p,q)\,v^2(r,s)\;X(N_1,m_1,i,p,r)\,X(N_2,m_2,j,q,s) \\
+ \dis\sum_{i+j=k,p+q=k,r+s=k} v^2(i,j)\,v^2(p,q)\,v^2(r,s)\;Y(N_1,m_1,i,p,r)\,Y(N_2,m_2,j,q,s) 
\earr \label{eq.pn18}
\ee 
where
\be
\barr{l}
X(N,m,k_1,k_2,k_3) = \dis\f{1}{\binom{N}{m}}\;\dis\sum_{\nu=0}^{min(m-k_1,k_2,m-k_3)}
\Lambda^\nu(N,m,k_1)\,\Lambda^\nu(N,m,m-k_2)\, \Lambda^\nu(N,m,k_3)\, d(N:\nu)\;,\\
Y(N,m,k_1,k_2,k_3) = \dis\f{1}{\binom{N}{m}^2} \dis\sum_{\nu_1=0}^{k_1} \dis\sum_{\nu2=0}^{k_2} \dis\sum_{\nu_3=0}^{m_1-k_3}  d(N:\nu_1) \,d(N:\nu_2) \\
\times \; \Lambda^{\nu_1}(N,m,m-k_1)\,\Lambda^{\nu_2}(N,m,m-k_2)\, \Lambda^{\nu_3}(N,m,k_3)\,\l|U(f_m \nu_1 f_m \nu_2\,;\, f_m \nu_3)\r|^2\;.
\earr \label{eq.pn18a}
\ee 
In summary, Eqs. (\ref{eq.pn7a}), (\ref{eq.pn11}), (\ref{eq.pn11a}), (\ref{eq.pn18}) and (\ref{eq.pn18a}) give the formulas for the second, fourth and sixth moment. In applying these formulas, the only unknown is $\l|U\r|^2 = \l|U(f_m \nu_1 f_m \nu_2\,;\, f_m \nu_3)\r|^2$. Exact formula for $\l|U\r|^2$ is not available but an approximate formula, in a special situation, follows from the asymptotic results described in the following Section 4. Before proceeding further, it is important to mention that our interest is obtaining the fourth reduced moment $\mu_4(m_1,m_2)$ and the reduced sixth moment $\mu_6(m_1,m_2)$,
\be
\mu_4(m_1,m_2) = 2 + q(m_1,m_2) = \dis\f{\overline{\lan H^4\ran^{m_1,m_2}}}{\l[\overline{\lan H^2\ran^{m_1,m_2}}\r]^2}\;,\;\;\; \mu_6(m_1,m_2) = \dis\f{\overline{\lan H^6\ran^{m_1,m_2}}}{\l[ \overline{\lan H^2\ran^{m_1,m_2}}\r]^3}\;.
\label{eq.pnm46}
\ee 
Then, in terms of $\overline{\lan H^2\ran^{m_1,m_2}}$ and the functions $X$, $Z$ and $Y$ defined above, we have,
\be
q(m_1,m_2) = \dis\f{\dis\sum_{i+j=k,p+q=k}
v^2(i,j)\,v^2(p,q)\;Z(N_1,m_1,i,p)\, Z(N_2,m_2,j,q)}{\l[\overline{\lan H^2\ran^{m_1,m_2}}\r]^2}\;,
\label{eq.pn19a}
\ee
and
\be
\barr{l}
\mu_6(m_1,m_2) = 5 + 6q(m_1,m_2) \\ \\
+ 3 \; \dis\f{\dis\sum_{i+j=k,p+q=k,r+s=k} v^2(i,j)\,v^2(p,q)\,v^2(r,s)\;X(N_1,m_1,i,p,r)\,X(N_2,m_2,j,q,s)}{\l[\overline{\lan H^2\ran^{m_1,m_2}}\r]^3} \\
+ \dis\f{\dis\sum_{i+j=k,p+q=k,r+s=k} v^2(i,j)\,v^2(p,q)\,v^2(r,s)\;Y(N_1,m_1,i,p,r)\,Y(N_2,m_2,j,q,s)}{\l[\overline{\lan H^2\ran^{m_1,m_2}}\r]^3} \;.
\earr \label{eq.pn20a}
\ee
Calculating $\mu_6(m_1,m_2)$ using Eqs. (\ref{eq.pn19a}) and  (\ref{eq.pn20a}) and comparing this with the result from Eq. (\ref{eq.q3a}), the $q$-normal form for the eigenvalue density is established in the next Section. 

\section{Asymptotic limit formulas and $q$-normal form of eigenvalue density}
\label{sec4}

\subsection{Asymptotic limit formulas}

In the dilute or asymptotic limit defined by $N_1,N_2 \rightarrow \infty$, $m_1,m_2 \rightarrow \infty$, $m_1/N_1\,m_2/N_2 \rightarrow 0$ and $k$ finite, it is easy to derive the formulas for the second, fourth and sixth moment by using the simple rules described first in \cite{MF} and also given in detail in \cite{Man-th} (see also Appendix in \cite{MK-1}). Firstly, it is easy to see that the second moment is given by
\be
\overline{\lan H^2\ran^{m_1,m_2}} = \dis\sum_{i+j=k} v^2(i,j)\, \binom{m_1}{i} \binom{N_1}{i} \binom{m_2}{j} \binom{N_2}{j}\;.
\label{eq.pn19}
\ee
For the fourth moment we need $Z(N,m,k_1,k_2)$ introduced in Eq. (\ref{eq.pn11}) and this is,
\be
Z(N,m,k_1,k_2) = \overline{\lan A(k_1)B(k_2)A(k_1)B(k_2)\ran^m}
\label{eq.pn20}
\ee
where in the $m$-particle average, the A and B are $H$ operators defining a EGUE($k_1$) and another independent EGUE($k_2$) and we use the same symbol for the binary correlated pairs. Also, note that in this derivation, the $m$ fermions are in $N$ single particle states giving matrix dimension $\binom{N}{m}$. The rule in \cite{MF} for commuting $B$ across $A$ gives $\lan ---- A(P)B(Q)A(P)---\ran^m = \f{\binom{m-Q}{P}}{\binom{m}{P}} \lan ---- A(P)A(P)B(Q)---\ran^m$. Now, combining this with Eq. (\ref{eq.pn19}) gives
\be
Z(N,m,k_1,k_2) = \binom{m-k_2}{k_1}\binom{m}{k_2} \binom{N}{k_1} \binom{N}{k_2}
\label{eq.pn21}
\ee
Proceeding further for the sixth moment we need $X(N,m,k_1,k_2,k_3)$ and $Y(N,m,k_1,k_2,k_3)$. Formula for $X$ s,
\be
\barr{rcl}
X(N,m,k_1,k_2,k_3) & = & \overline{\lan A(k_1)B(k_2)A(k_1)C(k_3)B(k_2)C(k_3)\ran^m} \\
& = & \dis\f{\binom{m-k_2}{k_1}}{\binom{m}{k_1}}\,\dis\f{\binom{m-k_2}{k_3}}{\binom{m}{k_3}}\, \overline{\lan [H(k_1)]^2\ran^{m}}\;\;\;\overline{\lan [H(k_2)]^2\ran^{m}}\;\;\;\overline{\lan [H(k_3)]^2\ran^{m}} \\
& = & \binom{m-k_2}{k_1}\,\binom{m-k_2}{k_3}\,\binom{m}{k_2}\, \binom{N}{k_1}\, \binom{N}{k_2}\, \binom{N}{k_3}\;.
\earr \label{eq.pn22}
\ee
Following the same rules as above, we have for $Y$,
\be
\barr{rcl}
Y(N,m,k_1,k_2,k_3) & = & \overline{\lan A(k_1)B(k_2)C(k_3)A(k_1)B(k_2)C(k_3)\ran^m} \\
& = & \dis\f{\binom{m-k_2-k_3}{k_1}}{\binom{m}{k_1}}\,\overline{\lan [H(k_1)]^2\ran^{m}}\;\;\;\overline{\lan B(k_2)C(k_3)B(k_2)C(k_3)\ran^{m}} \\
& = & \binom{m-k_2-k_3}{k_1}\,\binom{N}{k_1}\,Z(N,m,k_2,k_3) \\
& =& \binom{m-k_2-k_3}{k_1}\,\binom{m-k_3}{k_2}\,\binom{m}{k_3}\,\binom{N}{k_1}\,\binom{N}{k_2}\,\binom{N}{k_3}\;.
\earr \label{eq.pn23}
\ee
The asymptotic formulas given by Eqs. (\ref{eq.pn19}), (\ref{eq.pn21}), (\ref{eq.pn22}) and (\ref{eq.pn23}) also follow from the finite $N$ formulas given in Section 3 by noting that in the dilute limit,
\be
\Lambda^{k_1}(N,m,k_2) \stackrel{N \rightarrow \infty}{\longrightarrow}  \dis\binom{m-k_1}{k_2} \; \dis\binom{N}{k_2}\;,\;\;\; \dis\binom{N-p}{r} \stackrel{p/N \rightarrow 0}{\longrightarrow} 
\dis\frac{N^r}{r!}\;,\;\; d(N:\nu) \stackrel{\nu/N \rightarrow
0}{\longrightarrow} \dis\frac{N^{2\nu}}{ (\nu!)^2}\;.
\label{eq.pn24}
\ee
Using these, it is easy to derive the following:
\begin{itemize}

\item Putting $\nu=k_1$ in the finite $N$ formula for $Z(N,m,k_1, k_2)$ in Eq. (\ref{eq.pn11a}) will give the dilute limit formula given in Eq. (\ref{eq.pn21}).

\item Putting $\nu=k_2$ in the finite $N$ formula for
$X(N,m,k_1,k_2,k_3)$ in Eq. (\ref{eq.pn18a}) will give its dilute limit formula given in Eq. (\ref{eq.pn22}).

\item Finally,  putting $\nu_1=k_1$,  $\nu_2=k_2$ and $\nu_3=k_1+k_2$ in the finite $N$ formula for $Y(N,m,k_1,k_2,k_3)$ given by Eq. (\ref{eq.pn18a}) will simplify to the following,
\be
\barr{rcl}
Y(N,m,k_1,k_2,k_3) & = & \dis\binom{m-k_1-k_2}{k_3}\,\dis\binom{m}{k_1}\,\dis\binom{m}{k_2}\,\dis\binom{N}{k_1}\,\dis\binom{N}{k_2}\,\dis\binom{N}{k_3}\\ 
& \times & \l|U(f_m k_1 f_m k_2 ; f_m\, k_1+k_2)\r|^2\;.
\earr
\label{eq-y}
\ee
Comparing it with the dilute limit formula for $Y$ given in Eq. (\ref{eq.pn23}), will give for the $U$-coefficient,
\be
\l|U(f_m k_1 f_m k_2 ; f_m\, k_1+k_2)\r|^2= \l\{\binom{m}{k_1}\r\}^{-1}\,\binom{m-k_2}{k_1}\;.
\label{eq.pn25}
\ee
As described ahead, the asymptotic formula for $Y$ given by Eq. (\ref{eq.pn23}) is used in numerical calculations.

\end{itemize}

\subsection{Numerical results: $q$-normal form of the eigenvalue density}

\begin{figure}[h]
\centering
    \includegraphics[width=15.75cm,height=10.75cm]{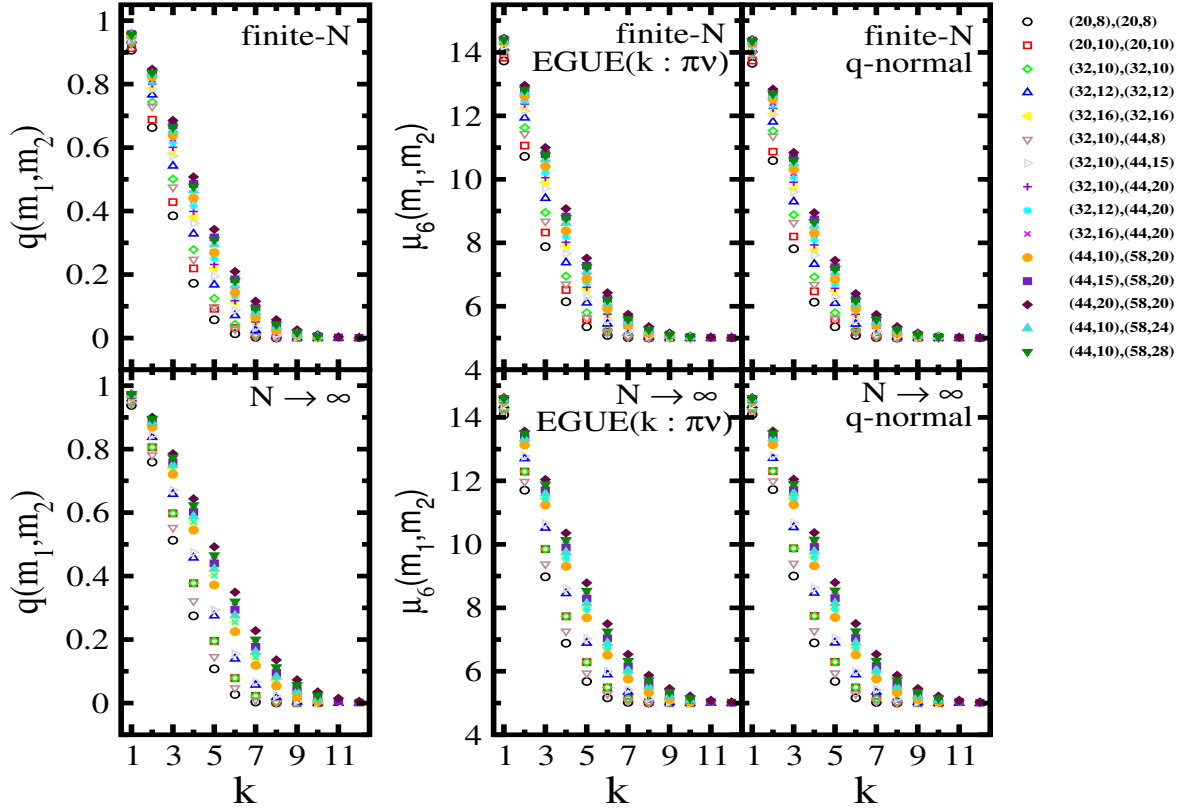} 
    \caption{Variation in $q(m_1,m_2)$ (left panel),  $\mu_6(m_1,m_2)$ for EGUE($k:\bpi\bnu$) (middle panel) and $\mu_6(m_1,m_2)$ obtained from $q$-normal (right panel) as a function of interaction rank $k$, for different values of $(N_1,m_1),(N_2,m_2)$ as indicated in the figure. The upper panel shows the results in finite-$N$ limit and the bottom panel shows the results in the asymptotic limit ($N \to \infty$).}
    \label{fig-1}
\end{figure}

In order to establish that the eigenvalue density generated by EGUE($k:\bpi\bnu$) is $q$-normal, we made several numerical calculations using the formulas given in Sections 3 and 4.1. Firstly, for simplicity and for making the results parameter free, we assume that $v^2(i,j) = v^2 = 1$ independent of $i$ and $j$. With this, both $q$ and $\mu_6$ will be independent of the parameter $v^2$.  For several different values of $(N_1,m_1),(N_2,m_2)$ and varying interaction rank $k=1$ to $k=min(m_1,m_2)$, we have calculated the value of $q(m_1,m_2)$ for finite-$N$ using Eqs. (\ref{eq.pn19a}), (\ref{eq.pn11a}) and (\ref{eq.pn7a}). Using this $q(m_1,m_2)$ value and Eq. (\ref{eq.pn20a}), calculated is $\mu_6(m_1,m_2)$. Here, for the function $X$ used is Eq. (\ref{eq.pn18a}). However, for calculating the last term in Eq. (\ref{eq.pn20a}), as the $U$-coefficient in Eq. (\ref{eq.pn18a}) is not available, we use the asymptotic limit formula for $Y$ given in Eq. \eqref{eq.pn23}; note that the last term $\sim q^3$ and hence small in general.  Results for $q$ value are shown in Figure \ref{fig-1}.  We also compare the $\mu_6$ value given by EGUE($k:\bpi\bnu$)  with the result from $q$-normal as given by (\ref{eq.q3a}) and the results are shown in Figure \ref{fig-1}.  The bottom panel in Figure \ref{fig-1} gives the corresponding results in the asymptotic limit (denoted by $N \to \infty$ in the figure).   Note that we have used $(N_1,N_2)$ values typical for heavy atomic nuclei; see \cite{KM-15}.  As can be seen from Figure \ref{fig-1}, the values of $q$ differ in the finite-$N$ and asymptotic limits.  It is seen that for sufficiently large values of $(m_1,m_2)$, the approach to semi-circle form (i.e. $q(m_1,m_2)=0$ or $\mu_6(m_1,m_2)=5$) is much faster as $k$ value increases and typically this happens for $k \sim min(m_1,m_2)/2$. However, in the approach to Gaussian (this needs $q(m_1,m_2)=1$ or $\mu_6(m_1,m_2)=15$) one sees significant departures even for $k=1$ and $2$.  In general, the asymptotic results differ from finite $N$ results by 10\%.  The difference in $\mu_6(m_1,m_2)$ from finite-$N$ formulas and from the $q$-normal form is in general less than 1\%, thus proving that the eigenvalue density generated by EGUE($k:\bpi\bnu$) is $q$-normal.

\begin{figure}[h]
\centering
    \includegraphics[width=15.75cm,height=10.75cm]{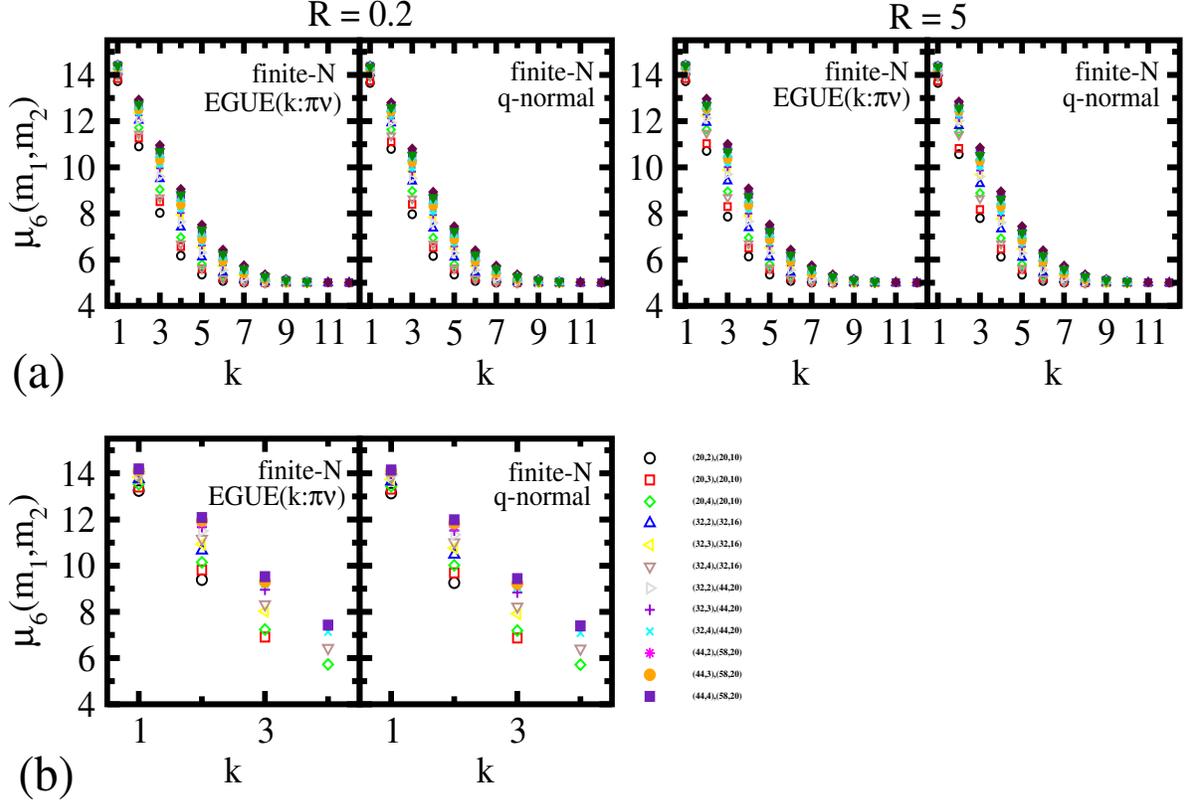} 
    \caption{Variation in $\mu_6(m_1,m_2)$ for EGUE($k:\bpi\bnu$) and $\mu_6(m_1,m_2)$ obtained from $q$-normal as a function of interaction rank $k$: (a) for $R = 0.2$ and $R = 5$ with different values of $(N_1,m_1),(N_2,m_2)$ as indicated in the Fig. \ref{fig-1}, (b) for small values of $m_1$ with different values of $(N_1,m_1),(N_2,m_2)$ as indicated in the figure.}
    \label{fig-1a}
\end{figure}

In order to further establish that the $q$-normal form is a general result, we have performed 
further calculations by considering (i) $v^2(i,j)$ not a constant and (ii) $m_1$ is very small and
$m_2$ is fairly large with $k$ changing from $k=1$ to $k=4$. Shown in Fig.  \ref{fig-1a}(a) are results obtained 
by using $v^2(i,j)=R\,v^2$ for $i,j >0$ and $v^2(i,j)= \,v^2$ for $i$ or $j = 0$. Then, the strength of the interaction $v^2(i,0)$ [or $v^2(0,j)$] acting in the $\bpi$ space (or in $\bnu$ space) differ by a factor $R$ from the strength of the interaction involving both $\bpi$ and $\bnu$ spaces. It is seen that in all the examples shown in the figure,
the difference in the value of $\mu_6(m_1,m_2)$ from finite-$N$ formulas and the $q$-normal form is less than 1\%. Thus, the difference in $v^2(i,j)$ strengths will not change the $q$-normal form. In a second set of
calculations, with results shown in Fig.  \ref{fig-1a}(b), used are $m_1=2$, $3$ or $4$ with $m_2$ large (results 
will be same if we interchange $m_1$ and $m_2$). Again we see that the difference in the value of $\mu_6(m_1,m_2)$ from finite-$N$ formulas and the $q$-normal form is less than 1\%. Thus, the $q$-normal form for the eigenvalue density is a general result for two species fermion systems. We will now turn to boson systems
in the next Section.

\section{Extension to two species boson systems}
\label{sec5}

Extension of EGUE($k:\bpi\bnu$) to two species boson systems is of considerable interest as this bosonic ensemble BEGUE($k:\bpi\bnu$) is relevant for two-component BEC systems \cite{BEC-1,BEC-2,BEC-3} and also for the proton-neutron interacting boson model of atomic nuclei \cite{IBA-1,IBA-2}.

Let us consider a system with two types of spinless bosons $\bpi$ and $\bnu$. Say this system consists of $m_1$ number of $\bpi$ bosons in $N_1$ number of single particle (sp) states and $m_2$ number of $\bnu$ bosons in $N_2$ number of sp states. This is
similar to protons ($\bpi$) and neutrons ($\bnu$) in interacting boson model of atomic nuclei. Further, the system Hamiltonian ($H$) operator is assumed to be $k$-body preserving $(m_1,m_2)$.  Thus, we have BEGUE($k:\bpi\bnu$) (‘B’ here stands for bosonic).  The embedding algebra for this system is again $U_\bpi(N_1) \oplus U_\bnu(N_2)$. 

One can derive the equations for $q$ and $\mu_6(m_1,m_2)$ for BEGUE($k:\bpi\bnu$) by applying the well known $N \rightarrow -N$ symmetry,  i.e.  in the fermion results replace $N$ by $-N$ and then take the absolute value of the final result; see \cite{Ko-05, KM-15} for
discussion on this property.  Firstly, it is easy to see that, for BEGUE($k$),  the $m$ boson space dimension $D_B(N,m)=\binom{N+m-1}{m}$ follows from the $m$ fermion space dimension $D_F(N,m)=\binom{N}{m}$ by replacing $N$ by $-N$ in $\binom{N}{m}$ and taking the absolute value giving in general,
\be
\binom{N+r}{s} \stackrel{N \rightarrow -N}{\longrightarrow} 
\dis\binom{N-r+s-1}{s}\;.
\label{eq.pn26}
\ee
As $m$ boson states should be symmetric under interchange of any two bosons, the  irrep $f_m^B =\{m\}$; the totally symmetric irrep of $U(N)$.  Moreover \cite{Ko-05},
\be
\barr{rcl}
\Lambda^\nu(N,m,k) \stackrel{bosons}{\longrightarrow}  \Lambda^\nu_B(N,m,k) 
& = & 
\l|\dis\binom{m-\nu}{k}\,\dis\binom{-N-m+k-\nu}{k}\r| \\
\\
& = & \dis\binom{m-\nu}{k}\,\dis\binom{N+m+\nu-1}{k} \;.
\earr \label{eq.pn27}
\ee
Also, for bosons the irreps $\nu$ for a $k$-body operator take the values
$\nu=0,1,\ldots,k$ as it is for fermions but for the fact that they correspond 
to the Young tableaux $\{2\nu,\nu^{N-2}\}$. Also, the $N \rightarrow -N$ 
symmetry will give 
\be
d(N:\nu) \stackrel{bosons}{\longrightarrow} d_B(N:\nu) ={\binom{N+\nu-1}{\nu}}^2-{\binom{N+\nu-2}{\nu-1}}^2\;.
\label{eq.pn28}
\ee

\begin{figure}[h]
\centering
    \includegraphics[width=15.75cm,height=5.275cm]{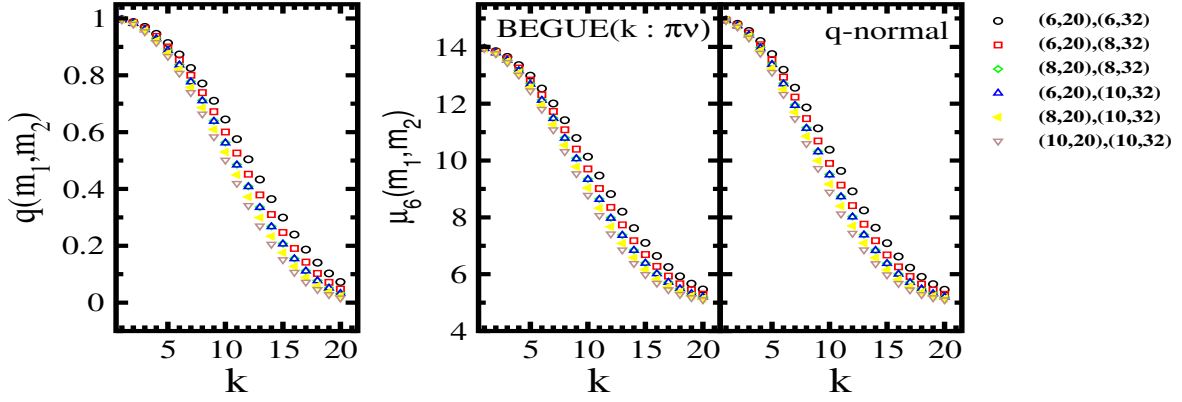} 
    \caption{Variation in $q(m_1,m_2)$ (left panel),  $\mu_6(m_1,m_2)$ for BEGUE($k:\bpi\bnu$) (middle panel) and $\mu_6(m_1,m_2)$ obtained from $q$-normal (right panel) as a function of interaction rank $k$, in finite-$N$ limit, for different values of $(N_1,m_1),(N_2,m_2)$ as indicated in the figure.  See text for further details.}
    \label{fig-2}
\end{figure}

For BEGUE($k:\bpi\bnu$), it is possible to apply the  $N \rightarrow -N$
symmetry but now for both $N_1$ and $N_2$ using Eq. (\ref{eq.pn26}).   Then,
$\binom{N_1+r}{s}$ will change to $\binom{N_1-r+s-1}{s}$ and $\binom{N_2+t}{u}$
changes to $\binom{N_2-t+u-1}{u}$. Similarly $\Lambda^{\nu}(N_1,m_1,k)$,
$d(N_1:\nu)$, $\Lambda^{\nu}(N_2,m_2,k)$ and $d(N_2:\nu)$ will change to the
bosonic $\Lambda_B$ and $d_B$ via Eqs. (\ref{eq.pn27}) and (\ref{eq.pn28}).
Using Eqs. (\ref{eq.pn26}), (\ref{eq.pn27}) and (\ref{eq.pn28}), it is
possible to write the following formulas,
\be
\barr{l}
Z_B(N,m,k_1,k_2) = \dis\f{1}{\binom{N+m-1}{m}}\;\dis\sum_{\nu=0}^{min(k_1,m-k_2)} \Lambda^\nu_B(N,m,m-k_1)\,\Lambda^\nu_B(N,m,k_2) \,d_B(N:\nu)\;,\\
X_B(N,m,k_1,k_2,k_3) = \dis\f{1}{\binom{N+m-1}{m}}\;\dis\sum_{\nu=0}^{min(m-k_1,k_2,m-k_3)}
\Lambda^\nu_B(N,m,k_1)\,\Lambda^\nu_B(N,m,m-k_2)\,  \\
\times \; \Lambda^\nu_B(N,m,k_3)\, d_B(N:\nu)\;,\\
Y_B(N,m,k_1,k_2,k_3) = \dis\f{1}{\binom{N+m-1}{m}^2} \dis\sum_{\nu_1=0}^{k_1} \dis\sum_{\nu2=0}^{k_2} \dis\sum_{\nu_3=0}^{m_1-k_3}  d_B(N:\nu_1) \,d_B(N:\nu_2) \\
\times \; \Lambda^{\nu_1}_B(N,m,m-k_1)\,\Lambda^{\nu_2}_B(N,m,m-k_2)\, \Lambda^{\nu_3}_B(N,m,k_3)\,\l|U(f^B_m \nu_1 f^B_m \nu_2\,;\, f^B_m \nu_3)\r|^2\;.
\earr \label{eq.pn18b}
\ee 
Using Eq. \eqref{eq.pn18b},  for several different values of $(N_1,m_1),(N_2,m_2)$ and varying interaction rank $k=1$ to $k=min(m_1,m_2)$, we have calculated the value of $q(m_1,m_2)$ in the finite-$N$ limit using Eqs. (\ref{eq.pn19a}), (\ref{eq.pn11a}) and (\ref{eq.pn7a}) for BEGUE($k:\bpi\bnu$). Then,  using this $q(m_1,m_2)$ value and Eq. (\ref{eq.pn20a}), calculated is $\mu_6(m_1,m_2)$.  As the formula for Racah coefficient is not available, we assume that the contribution from the $Y$ terms is negligible. Note that the last term $\sim q^3$ and hence is small in general. Its contribution is less than 1\% as verified numerically in the asymptotic limit for EGUE($k:\bpi\bnu$). Results are shown in Fig. \ref{fig-2}.  The values of $\mu_6(m_1,m_2)$ for BEGUE($k:\bpi\bnu$) are quite close to that obtained using the result from $q$-normal for various combinations of $(N_1,m_1),(N_2,m_2)$, thus proving that the eigenvalue density generated by BEGUE($k:\bpi\bnu$) is also $q$-normal.

\section{Conclusions}
\label{sec6}

Embedded ensembles have been used successfully to understand the many-body eigenvalue densities (and other properties) of complex finite quantum systems; see \cite{Ko-book} and references therein.  In this paper, we have established that the two species $k$-body embedded Gaussian unitary ensembles generate $q$-normal form of the eigenvalue density.  Assuming $k$-body system Hamiltonian that preserves number of particles (fermions or bosons) $(m_1,m_2)$, we derived the formulas for parameter $q(m_1,m_2)$ [that is related to the fourth moment $\mu_4(m_1,m_2)$] and the sixth moment $\mu_6(m_1,m_2)$.  We have derived exact group theoretical formulas for the fourth and sixth order moments for both EGUE($k:\bpi\bnu$) and BEGUE($k:\bpi\bnu$) [respectively corresponding to fermionic and bosonic embedded Gaussian unitary ensembles of $k$-body interactions with two species $\bpi$ and $\bnu$].  The corresponding asymptotic results are presented in Section \ref{sec4} for EGUE($k:\bpi\bnu$).  Numerical results in Figs. \ref{fig-1} and \ref{fig-1a} for EGUE($k:\bpi\bnu$) and in Fig. \ref{fig-2} for BEGUE($k:\bpi\bnu$), clearly show that the eigenvalue densities are $q$-normal.  In order to further establish that the $q$-normal form is a general result, we have performed calculations by considering (i) $v^2(i,j)$ not a constant and (ii) $m_1$ is very small and
$m_2$ is fairly large with changing rank of interactions $k$.  Varying $v^2(i,j)$ does not change the $q$-normal form of the eigenvalue density. This may be similar to the fact that the eigenvalue density for Wigner matrices with an inhomogeneous variance is still semi-circle in the large matrix limit \cite{link-notes}. As the fourth moment gives $q(m_1,m_2)$ and this defines the $q$-normal form, to establish if the $q$-normal form for eigenvalue density is good, we have verified the closeness of the sixth moment $\mu_6(m_1,m_2)$ generated by the ensemble with that generated by the $q$-normal form. In fact, a further check will come by comparing the eight moment $\mu_8(m_1,m_2)$ but here the ensemble formula (not given in this paper) is much more complex involving $U$-coefficients for which no (exact/approximate) analytical formulas or numerical methods are available.
In future, it will be interesting to consider various other modified fermionic and bosonic EE (see \cite{Ko-book,KC-1,KC-2} for some examples) and examine the form of the eigenvalue density generated by them vis-a-vis the $q$-normal form.

\acknowledgments

M. V.  acknowledges financial support from CONACYT project Fronteras 10872.

\ed